\documentclass[12pt]{article}
\usepackage{epubtk}
\usepackage{amssymb}
\usepackage{amsmath}
\usepackage{graphicx}
\usepackage[tiny,bf]{titlesec}
\usepackage{hyperref}
\usepackage{mciteplus}

\begin{document}

\newcommand{\simpropto}{
\begin{array}{c}
\propto \\[-1.7ex] \sim
\end{array}}

\newcommand{\dd}{d}
\newcommand{\DD}{D}
\newcommand{\im}{i}
\newcommand{\ee}{e}
\newcommand{\perpperp}{\perp\!\!\perp}
\newcommand{\nn}{\nonumber\\}
\newcommand{\unit}[1]{\, {\rm #1}}

\newcommand{\Msun}{{\rm M}_\odot}
\newcommand{\vel}{\textsl{v}}
\newcommand{\inn}{{\rm in}}
\newcommand{\out}{{\rm ou}}
\newcommand{\abh}{a_\bullet}
\newcommand{\Mbh}{M_\bullet}
\newcommand{\Mbhdot}{\dot{M}_\bullet}
\newcommand{\Qbh}{Q_\bullet}

\newcommand{\bg}{\bm{g}}
\newcommand{\bp}{\bm{p}}
\newcommand{\bx}{\bm{x}}
\newcommand{\smalltc}{{\scriptscriptstyle T}}

\newcommand{\plus}{{\scriptscriptstyle +}}
\newcommand{\minus}{{\scriptscriptstyle -}}
\newcommand{\plusminus}{{\scriptscriptstyle \pm}}
\newcommand{\minusplus}{{\scriptscriptstyle \mp}}
\newcommand{\zero}{{\scriptstyle 0}}
\newcommand{\one}{{\scriptstyle 1}}
\newcommand{\two}{{\scriptstyle 2}}
\newcommand{\three}{{\scriptstyle 3}}
\newcommand{\comma}{{\scriptscriptstyle ,}}
\newcommand{\smallzero}{{\scriptscriptstyle 0}}
\newcommand{\smallone}{{\scriptscriptstyle 1}}
\newcommand{\Cz}{{\tilde C}}
\newcommand{\nY}[1]{{}_{#1\,} \! Y}

\hyphenpenalty=3000

\newcommand{\trajectoryunstablefig}{
    \begin{figure}[b!]
    \begin{center}
    \leavevmode
    \includegraphics[scale=.5]{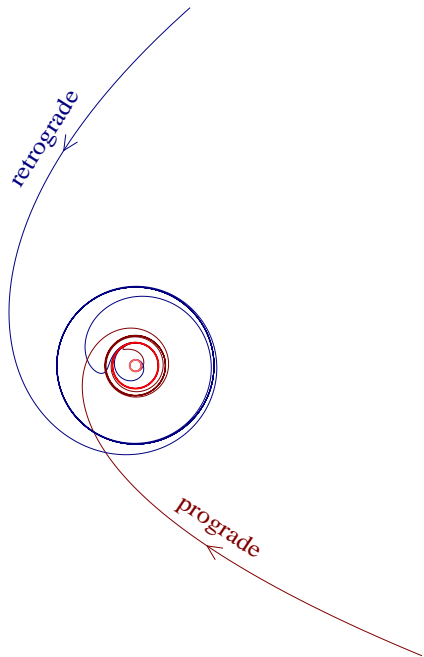}
    \caption[1]{
    \label{trajectoryunstable}
\small
Prograde and retrograde streams accreted from outside the horizon
of a rotating black hole
focus respectively along the outgoing and ingoing principal null directions
as they approach the inner horizon.
The accreted streams fuel ongoing inflation at (just above)
the inner horizon.
The (red) circles show the outer and inner horizons of the black hole,
which here has spin $a = 0.8 M$.
    }
    \end{center}
    \end{figure}
}

\newcommand{\penrosekerrinflationfig}{
    \begin{figure}[t]
    \begin{center}
    \leavevmode
    \includegraphics[scale=.9]{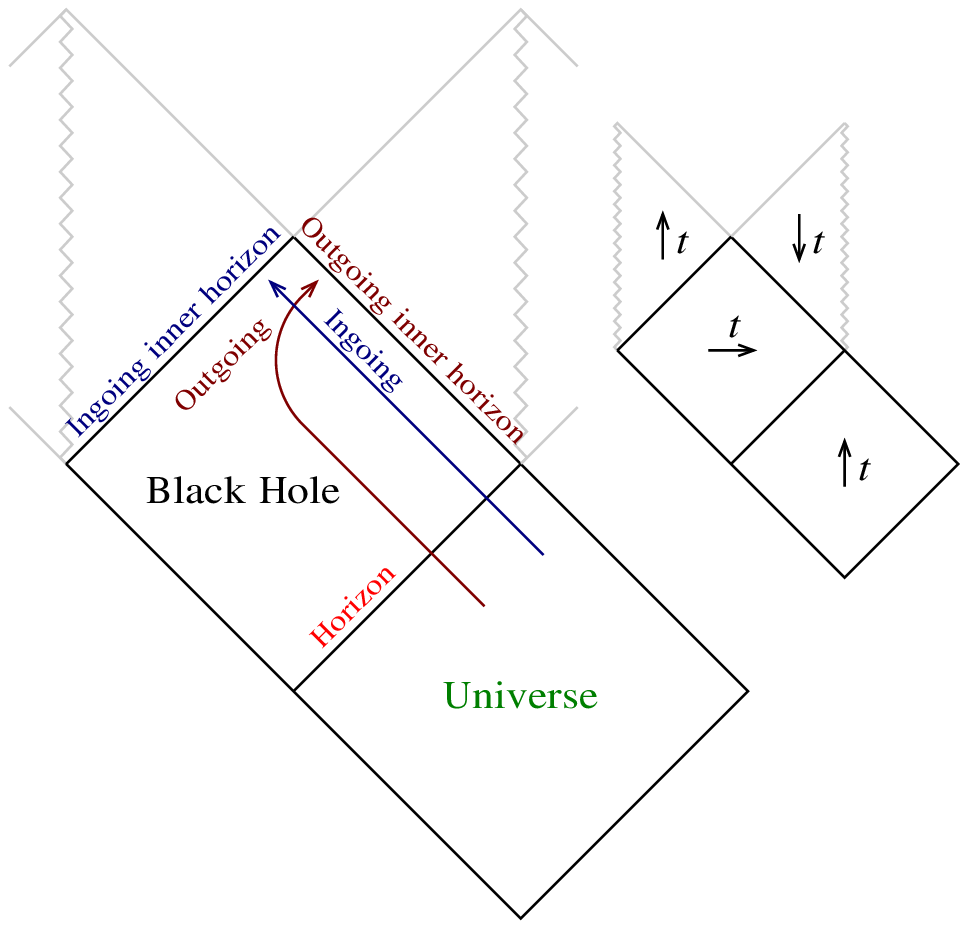}
    \caption[1]{
    \label{penrosekerrinflation}
\small
Partial Penrose diagram illustrating why the Kerr geometry
is subject to the inflationary instability.
The inset shows the direction of coordinate time $t$
in the various regions
(proper time of course always increases upward
in a Penrose diagram).
The inflationary instability destroys the inner horizon,
so the (light grey) Kerr regions beyond the inner horizon
do not exist in a real black hole.
    }
    \end{center}
    \end{figure}
}

\markboth{A. J. S. Hamilton}
{The Black Hole Particle Accelerator}

\title{\bf\large The Black Hole Particle Accelerator \\
as a Machine to make Baby Universes}

\author{\normalsize
A. J. S. Hamilton \\
\small
JILA, Box 440, U. Colorado, Boulder, CO 80309, USA \\
\small
Andrew.Hamilton@colorado.edu
}

\date{\small\today}

\maketitle

\begin{abstract}
General relativity predicts that
the inner horizon of an astronomically realistic rotating black hole
is subject to the mass inflation instability.
The inflationary instability acts like a gravity-powered
particle accelerator of extraordinary power,
accelerating accreted streams of particles
along the principal outgoing and ingoing null directions
at the inner horizon
to collision energies that would, if nothing intervened,
typically exceed exponentially the Planck energy.
The inflationary instability is fueled by ongoing accretion,
and is occurring inevitably in essentially every black hole in our Universe.
This extravagant machine,
the Black Hole Particle Accelerator, has the hallmarks
of a device to make baby universes.
Since collisions are most numerous inside supermassive black holes,
reproductive efficiency requires our Universe
to make supermassive black holes efficiently,
as is observed.
\end{abstract}

\begin{center}
\it
Essay written for the Gravity Research Foundation 2013 Awards \\
for Essays on Gravitation.
\end{center}

\section*{Introduction}

Smolin
\cite{Smolin:1992,Smolin:1997,Smolin:2013}
has advocated the idea of cosmological natural selection.
Suppose that, among the sets of laws of physics in the multiverse,
there are some that allow a universe to reproduce,
and to pass on that ability to its offspring.
Universes that can reproduce efficiently
can lead to a population explosion,
crowding the multiverse with their progeny.
If there exist laws of physics that allow for efficient reproduction,
then it would not be surprising if our Universe could reproduce.

Smolin
\cite{Smolin:1992,Smolin:1994}
proposed that the natural place for reproduction to occur is
at the singularities of black holes.
He suggested that quantum effects would replace
the singularity of each black hole by a bounce,
so that each black hole would effectively produce one baby universe.
If so,
then the most abundant universes would be those
that produce the largest number of black holes.

Recent theoretical research
\cite{Hamilton:2008zz,Hamilton:2010a,*Hamilton:2010b,*Hamilton:2010c,Hamilton:2011vr}
shows that, if general relativity is correct,
then rotating black holes inevitably generate enormous
(Planckian or hyper-Planckian) energies
in an ingenious machine at the inner horizon
that I call the Black Hole Particle Accelerator (BHPA).
If our Universe reproduces,
then the BHPA, not a singularity, surely provides the mechanism.

\section*{The Black Hole Particle Accelerator}

The Kerr \cite{Kerr:2007dk} geometry for a rotating black hole has,
unlike the Schwarzschild geometry,
not only an outer horizon but also an inner horizon.
The inner horizon is the boundary of predictability
(a Cauchy horizon),
a gateway to fun but bizarre pathologies,
such as wormholes, white holes, naked (timelike) singularities,
and closed timelike loops
\cite{Carter:1968a}.

\trajectoryunstablefig

Linear perturbation theory
shows that waves incident on the inner horizon amplify without bound,
suggesting instability
\cite{Chandrasekhar:1982},
but it was not until the seminal work of
Poisson \& Israel
\cite{Poisson:1990eh,*Barrabes:1990}
that the nonlinear development of the instability at the inner horizon
started to become clear.
\cite{Poisson:1990eh,*Barrabes:1990}
argued that counter-streaming between outgoing and ingoing streams
just above the inner horizon would cause the interior mass
to grow exponentially,
a phenomenon they dubbed ``mass inflation.''

\cite{Poisson:1990eh,*Barrabes:1990}
proposed that outgoing and ingoing streams would be
produced by Price tails of radiation generated during the collapse
of a black hole.
However,
the energy density of the least rapidly decaying (quadrupole) mode
of a Price tail decays as $t^{- 12}$,
falling below cosmic microwave background density
in about a second, for a stellar-mass black hole.
Thus the astronomically realistic situation is that outgoing
and ingoing streams are fed by ongoing accretion,
which quickly overwhelms any Price tail.

\penrosekerrinflationfig

Accretion, most probably of baryons and dark matter,
naturally produces both outgoing and ingoing streams.
Figure~\ref{trajectoryunstable}
illustrates a pair of example geodesics of (dark matter, say)
particles that free-fall from zero velocity at infinity.
A freely-falling particle that is prograde,
though necessarily ingoing at the outer horizon,
generically switches to being outgoing at the inner horizon
(physically, its angular momentum gives it an outward centrifugal push).
Conversely, retrograde particles generically remain ingoing
down to the inner horizon.

Figure~\ref{penrosekerrinflation} shows a partial Penrose diagram
that illustrates why the Kerr geometry
is subject to the inflationary instability.
The problem is that, to fall through the inner horizon,
outgoing and ingoing streams must fall through separate
outgoing and ingoing inner horizons into causally separated regions
where the timelike Boyer-Linquist time coordinate $t$
goes in opposite directions.
This requires that outgoing and ingoing streams exceed the speed
of light relative to each other, which is physically impossible.
In practice,
regardless of their initial orbital parameters,
outgoing and ingoing streams approaching the inner horizon
become highly focused along
the outgoing and ingoing principal null directions,
and enormously blueshifted relative to each other
\cite{Hamilton:2010a,*Hamilton:2010b,*Hamilton:2010c,Hamilton:2011vr}.
However tiny the initial streams might be,
the proper streaming energy density grows large enough
to become a source of gravity that competes with the native gravity
of the black hole.
In a fashion that is peculiarly general relativistic,
the gravitational force is in opposite directions
for the two streams,
accelerating them even faster through each other.
The streams accelerate exponentially,
powered by the gravity generated by their own counter-streaming.
This is inflation.

As Figure~\ref{penrosekerrinflation} illustrates,
outgoing particles stream through ingoing particles accreted in the future,
while
ingoing particles stream through outgoing particles accreted in the past.
The blueshift between the outgoing and ingoing streams increases exponentially.
Each stream sees approximately one black hole crossing time
elapse on the opposing stream for each $\ee$-fold increase of blueshift
\cite{Hamilton:2008zz}.
The center-of-mass collision energy between particles of rest mass
$\sim 1$--$10^3 \unit{GeV}$
reaches the Planck energy of $\sim 10^{19} \unit{GeV}$
after of order 100 $\ee$-folds.
This happens before the streaming energy density and curvature
hit the Planck density,
which takes of order 300 $\ee$-folds
\cite{Hamilton:2008zz}.
A few hundred $\ee$-folds corresponds to less than a second
for a stellar mass black hole,
or hours to years for a supermassive black hole.
Thus what happens during inflation at any time depends
on accretion during the immediate past and future,
but not on the history over cosmological timescales.

The enormous gravitational acceleration that drives inflation
eventually leads to collapse in the transverse directions
\cite{Hamilton:2008zz,Hamilton:2010a,*Hamilton:2010b,*Hamilton:2010c,Hamilton:2011vr}.
Collapse occurs after approximately $1/\dot{M}$ $\ee$-folds,
where $\dot{M}$ is the dimensionless ($c = G = 1$) accretion rate.
Thus the larger the accretion rate, the more rapidly collapse occurs.
If the accretion rate is larger than $\dot{M} \gtrsim 0.01$,
then collapse occurs before streams can blueshift
the $100$ $\ee$-folds needed to reach the Planck scale,
and presumably no baby universe can be made.
However,
accretion rates as high as $0.01$
occur only rarely,
such as during the initial collapse of a black hole,
or during a black hole merger,
or during occasional events of high accretion.
Astronomical black holes
spend the vast majority of their lives accreting more slowly,
therefore producing collisions at Planck energies and above.

\section*{Making baby universes}

Making baby universes is not easy
\cite{Berezin:1985,*Berezin:1985re,Farhi:1986ty,Farhi:1989yr,Linde:1991sk,Aguirre:2005nt,Aguirre:2007gy}.
The fundamental problem is that
making a baby universe appears to violate the second law of thermodynamics.
If a baby is a causal product of its mother,
then there are observers whose worldlines pass from mother to baby,
and such observers see a transition from a state of high entropy,
our Universe,
to one of low entropy,
a baby universe in an inflating, vacuum-dominated state
smooth over its horizon scale.

The BHPA has at least two ingredients essential for reproduction.
The first is that it drives energies up to the Planck scale
(and beyond).
If the genes of a universe are encoded for example
in the structure of its Calabi-Yau manifold,
then some mechanism is needed to drive the system to a sufficient
energy to allow a slight rearrangement of the Calabi-Yau.

The second ingredient is a mechanism to overcome the entropy barrier.
Our Universe apparently solves the entropy problem
at least in part by brute force:
every day,
it is carrying out prodigious numbers of collision experiments over a broad
range of Planckian and hyper-Planckian center-of-mass energies
in large numbers of black holes
throughout the Universe.

The BHPA has another feature
that it surely harnesses
to overcome the entropy barrier.
The most common way that systems in Nature reduce their
entropy is not by happenstance,
but rather by exploiting the free energy of a system out of
thermodynamic equilibrium.
The BHPA is far out of thermodynamic equilibrium:
initially,
it comprises two cold streams accelerating through each other
to enormous energies.

\section*{Supermassive black holes}

If babies are made where the largest number of collisions are taking place,
then supermassive black holes are implicated as the most likely mothers.
The number of collisions over any $\ee$-fold of energy is proportional
to $M^3$, the black hole mass cubed.
Two powers of $M$ come from collisions being a two-body process,
and a third power comes from the fact that the time spent
in any $\ee$-fold range of collision energy is proportional to the black hole
crossing time.

The hypothesis of cosmological natural selection
predicts that the most abundant universes in the multiverse
are those that reproduce most efficiently
\cite{Smolin:1992,Smolin:1994}.
If reproduction works as proposed in this essay,
then the most successful reproducers are those that make large numbers
of supermassive black holes.

In fact our Universe is remarkably
efficient at making supermassive black holes.
Observations indicate that most, if not all, massive galaxies
host supermassive black holes at their centers
\cite{Magorrian:1998}.
The existence of $\sim 10^9 \unit{\Msun}$ black holes
already at redshift $z \sim 6$
means that supermassive black holes were able to form
so fast as to pose a challenge to conventional models
that assume Eddington-limited growth by accretion~\cite{Begelman:2006db}.
As our Universe grows old,
almost all the matter in it may end up falling into
a supermassive black hole.

\section*{Challenge}

Are there laws of physics that allow universes to reproduce?
If so,
then chances are our Universe is using them in the BHPA.
Does reproduction work only for special sets of laws of physics?
Excellent.
Undoubtedly
those laws hold for our Universe.
Nature is offering clues.
Let us listen.

\bibliography{bh}

\end{document}